\documentclass[aps,amsmath,amssymb,twocolumn,prl]{revtex4-2}

\usepackage{graphicx}
\usepackage{dcolumn}
\usepackage{bm}
\usepackage{color}
\usepackage[normalem]{ulem}
\usepackage{amsmath}
\usepackage{braket}
\usepackage{hyperref}

\begin{document}
	\pagenumbering{arabic}
	
\title{Non-hermitian topology in multiterminal superconducting junctions}



\author{David Christian Ohnmacht}
\email{david.ohnmacht@uni-konstanz.de}
\thanks{These authors contributed equally}

\author{Valentin Wilhelm}
\thanks{These authors contributed equally}
\author{Hannes Weisbrich}
\author{Wolfgang Belzig}
\email{wolfgang.belzig@uni-konstanz.de}

\affiliation{Fachbereich Physik, Universit{\"a}t Konstanz, D-78457 Konstanz, Germany}
	
\begin{abstract}
Recent experimental advancements in dissipation control have yielded significant insights into non-hermitian Hamiltonians for open quantum systems. Of particular interest are the topological characteristics exhibited by these non-hermitian systems, that arise from exceptional points—distinct degeneracies unique to such systems.
In this study, we focus on Andreev bound states in multiterminal Josephson junctions with non-hermiticity induced by normal metal or ferromagnetic leads. By investigating several systems of different synthetic dimensions and symmetries, we predict fragile and stable non-hermitian topological phases in these engineered superconducting systems.
	\end{abstract}

\date{\today}

\maketitle

\emph{Introduction}.-- In the field of condensed matter physics, the incorporation of topology over the past few decades has significantly transformed our understanding of materials. This approach has paved the way for exploring exotic electronic states and unprecedented material properties, as evidenced for example by the emergence of topological insulators \cite{hasan2010colloquium} and topological superconductors \cite{sato2017topological}. These remarkable advancements have not only reshaped the landscape of condensed matter research but also sparked interest in connected fields, including the utilization of topologically protected quantum states for achieving robust quantum computation \cite{Nayak_Simon_Stern_Freedman_Sarma_2008,deng2016majorana,aasen2016milestones,karzig2017scalable}.

Over the past few years, the concept of synthetic topological systems has gained significant attention. Examples include topological photonics \cite{lu2014topological,wang2009observation,khanikaev2013photonic}, topological-driven Floquet systems \cite{rudner2020band,kitagawa2010topological}, topological electrical circuits \cite{imhof2018topolectrical}, topological multiterminal Josephson junctions \cite{riwar2016multi,eriksson2017topological,xie2017topological,xie2018weyl,deb2018josephson,xie2019topological,gavensky2019topological,houzet2019majorana,klees2020microwave,klees2021ground,Weisbrichtensor,PhysRevB.107.165301,Peralta_Gavensky_2023,riwar2019fractional,PhysRevB.107.035408}, and topological superconducting circuits  \cite{fatemi2021weyl,peyruchat2021transconductance,herrig2022cooper}. These engineered systems incorporate topological features in internal degrees of freedom, allowing the construction of topology beyond the inherent dimensionality of the system itself  \cite{weisbrich2021second,xie2022non}. Notably, topological signatures emerge within these systems, such as the prediction of a quantized transconductance in multiterminal Josephson junctions  \cite{riwar2016multi}, akin to the behavior observed in quantum Hall systems \cite{klitzing1980new,TKKN,hatsugai1993chern}.

Recent studies have expanded the exploration of topological concepts beyond isolated systems described by hermitian models. Instead, the focus has shifted towards open and dissipative setups that can be characterized by non-hermitian Hamiltonians. Surprisingly, non-hermitian systems have been found to exhibit novel topological phases not observed in their hermitian counterparts \cite{martinez2018topological,gong2018topological,kawabata2019symmetry,bergholtz2021exceptional,PhysRevLett.123.066405,Okuma2020,Manna2023a}, spanning diverse domains from classical metamaterials to condensed matter systems.

The intriguing topological properties of non-hermitian systems arise from a unique type of degeneracy called exceptional point (EP) \cite{dembowski2001experimental,doppler2016dynamically}, where eigenvalues and eigenvectors coalesce. These EPs and respective topologies have already demonstrated a wide range of remarkable phenomena, including enhanced sensing capabilities \cite{chen2017exceptional}, unidirectional lasing \cite{peng2016chiral}, and the presence of bulk Fermi arcs \cite{zhou2018observation}, none of which have analogous counterparts in hermitian systems. Non-hermitian EPs have recently emerged as a captivating frontier in the investigation of superconducting systems, focusing on the supercurrent characteristic and how it is related to the complex Andreev bound states (ABSs) spectrum \cite{cayao2023bulk, PhysRevLett.133.086301, PhysRevB.110.L201403, PhysRevB.109.214514,10.1063/5.0215522,Pino2024, cayao2024nonhermitianminimalkitaevchains, cayao2024nonhermitianmultiterminalphasebiasedjosephson}. In particular, it was shown in Ref.~\cite{PhysRevLett.133.086301} that EPs leave signatures in the current susceptibility. Other means to detect EPs could involve workarounds to circumvent the breakdown of adiabatic state transfer \cite{Uzdin_2011}, which has been demonstrated in experiment \cite{Doppler2016} and theory \cite{PhysRevLett.133.113802}. Furthermore, there has been an ongoing effort to study the appearance of Majorana bound states in hybrid structures in the context of non-hermiticity and EPs \cite{San-Jose2016,Avila2019,Okuma2019,Arouca2023} and dissipation \cite{Liu2022,Zhang2022}.

In this manuscript, we demonstrate that multiterminal Josephson junctions provide ideal platforms for engineered non-hermitian topology and EPs in different synthetic dimensions. Our work provides insight into the stability, topology and potential usability of these EPs. First, we analyze a two-terminal hybrid system where we find that EPs are fragile. By adding another dimension in the form of an additional superconducting terminal, we find stable EPs and windings ensuring topological stability. Lastly, we show that Weyl points arising in multiterminal Josephson junctions form exceptional rings \cite{bergholtz2021exceptional} for arbitrarily small non-hermiticity, but only when the spin-rotation symmetry is broken. In particular, we show that ferromagnets or spin-orbit coupling can be utilized as the source of broken spin-rotation symmetry. All of these observations can be explained in the language of the 38-fold way \cite{kawabata2019symmetry} when the superconducting phases are treated as synthetic dimensions. 



\begin{figure*}
\includegraphics[width=1\textwidth]{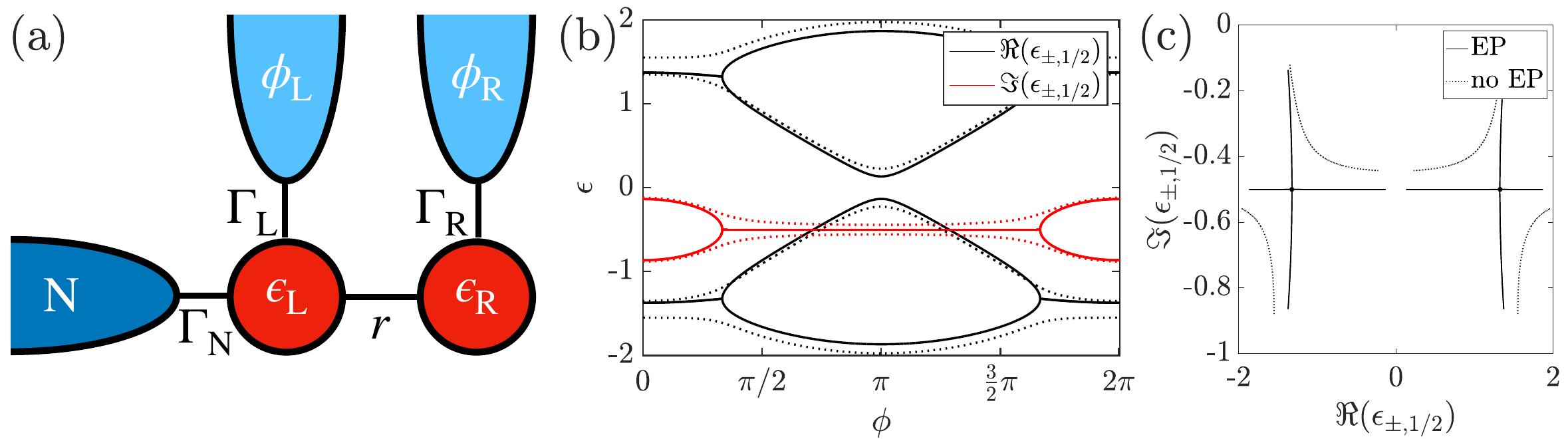}
\caption{(a) Schematic of a double quantum dot two-terminal junction where a normal lead adds a non-hermiticity in one of the dots. (b) Real- and imaginary part of the eigenvalues of the effective non-hermitian Hamiltonian, $\epsilon_{\pm,1/2}$ as functions of phase difference $\phi$.  Solid lines are for the symmetric case where EPs emerge (Parameters: $\Gamma_{\rm L/R} = r = 1$, $\Gamma_{\rm N} = 1$, $\epsilon_{\rm L/R} = 0$). Dotted lines are for the non-symmetric case where there are no EPs (Parameters: $\Gamma_{\rm R} = 1$, $\Gamma_{\rm L} = 1.2$, $\Gamma_{\rm N}= 1$, $r = 1$, $\epsilon_{\rm L/R} = 0$). (c) Energy spectrum $\epsilon_{\pm,1/2}(\phi)$ in the complex plane for both parameter configurations with EP and no EP shown in panel (b).}
\label{Fig1}
\end{figure*}

\emph{Fragile exceptional points in two-terminal superconducting junctions}.-- In the following we consider a double quantum dot junction that is tunnel coupled to two superconducting leads and a single normal lead, as depicted in Fig.~\ref{Fig1}(a). Models of this nature have been extensively studied in the past few decades in various different contexts \cite{martin2011josephson,recher2001andreev,governale2008real,pala2007nonequilibrium,eldridge2010superconducting,futterer2009nonlocal,herrmann2010carbon,tanaka2010correlated,tanaka2008andreev,jonckheere2009nonequilibrium,de2010hybrid}.
In the following, all relevant parameters are dimensionless and only fractions of parameters are of physical relevance. For small energies in the limit of $E\ll\Delta$ the system can be described by an effective Hamiltonian 
\begin{equation}
H_{\text{eff}}=\bm{d}^\dagger\begin{pmatrix}
\epsilon_\text{L}\tau_3 + \Gamma_{\rm L}\tau_1 -i\Gamma_{\text{N}}\tau_0 &r\tau_3\\
r\tau_3 & \epsilon_{\rm R}\tau_3 + \Gamma_{\rm R}e^{i\phi\tau_3}\tau_1
\end{pmatrix}\bm{d},\label{eq:eff}
\end{equation}
with the Nambu spinor $\bm{d}^\dagger=(d_{\text{L}\uparrow}^\dagger,d_{\text{L}\downarrow},d_{\text{R}\uparrow}^\dagger, d_{\text{R}\downarrow})$, the Pauli matrices in Nambu space $\tau_i$ ($i = 0,1,2,3$), the superconducting phase difference $\phi=\phi_\text{R}-\phi_\text{L}$, the dot energies $\epsilon_{\rm L/R}$, the effective coupling between the dots and the superconducting leads $\Gamma_{\text{L/R}}$ and between the left dot and the normal lead $\Gamma_\text{N}$. The coupling to the normal lead effectively introduces a non-hermitian term in the low energy Hamiltonian, resulting in a broadening of the superconducting bound states. The use of an effective non-hermitian Hamiltonian can be justified at sufficiently short timescales when jump terms in the Lindblad master equation are negligible \cite{bergholtz2021exceptional,PhysRevB.109.214514}. In the case of multiterminal Josephson junctions, the couplings to the normal lead and superconductors must be small compared to the superconducting gap $\Delta$. For the normal metal, this results in long lifetimes $\propto 1/\Gamma_{\rm N}$ of the bound states, and small couplings to superconducting leads results in a weak energy dependence of the self energies. In Ref.~\cite{SM}, we demonstrate how 
the effective Hamiltonian is derived from a Green's function approach in the large gap limit and moreover, show that the current obtained in this limit agrees well with the full current. In particular, in the large gap limit, we retrieve the same current formula as in Ref.~\cite{PhysRevLett.133.086301} using an action approach \cite{Ohnmacht2023}, meaning that the complex spectrum is directly related to the measurable current. Consequently, non-hermitian effects which are enabled by arbitrarily small non-hermiticities are most promising with respect to experimental implementation.

The spectrum consists of four complex ABSs, two at energies below (real) zero $\epsilon_{-,1/2}$ and two above zero $\epsilon_{+,1/2}$. In the case when the junction is symmetric ($\Gamma_{\rm L} = \Gamma_{\rm R}$ and $\epsilon_{\rm R} = \epsilon_{\rm L}$) EPs, where two ABSs energies coalesce, emerge for non-zero energies like it is seen in Fig.~\ref{Fig1}(b) (solid lines). In the symmetric case, the system is analogous to the minimal model described in Ref.~\cite{PhysRevB.110.L201403}. However, these EPs vanish when the aforementioned symmetry is broken, which is shown in Fig.~\ref{Fig1}(b) by the dotted lines. In this model, the symmetry enforces degeneracies of the two ABSs in the hermititan case, when the normal metal coupling is neglected. As soon as this degeneracy is lifted, no EPs can emerge. Because of this fragility, such EPs aren't suitable candidates for experimental identification.

This fragility can be explained by classifying the system in the language of the 38-fold way \cite{kawabata2019symmetry}. Note, that the EPs do not appear at zero energy. In order to classify them properly, one EP has to be shifted towards zero energy which breaks particle-hole symmetry (PHS). In addition, fixing a phase difference further breaks time reversal symmetry (TRS) of the junction \cite{riwar2016multi} rendering the system without any symmetries, which results in the system belonging to the AZ class A. The spectrum of the symmetric/asymmetric system is shown in Fig.~\ref{Fig1}(c), with a corresponding point/line gap. For the point gap present in our model the EP marks no topological transition as there is no nontrivial invariant for a point gap in the zero-dimensional case \cite{kawabata2019symmetry}.

\begin{figure*}
\includegraphics[width=1\textwidth]{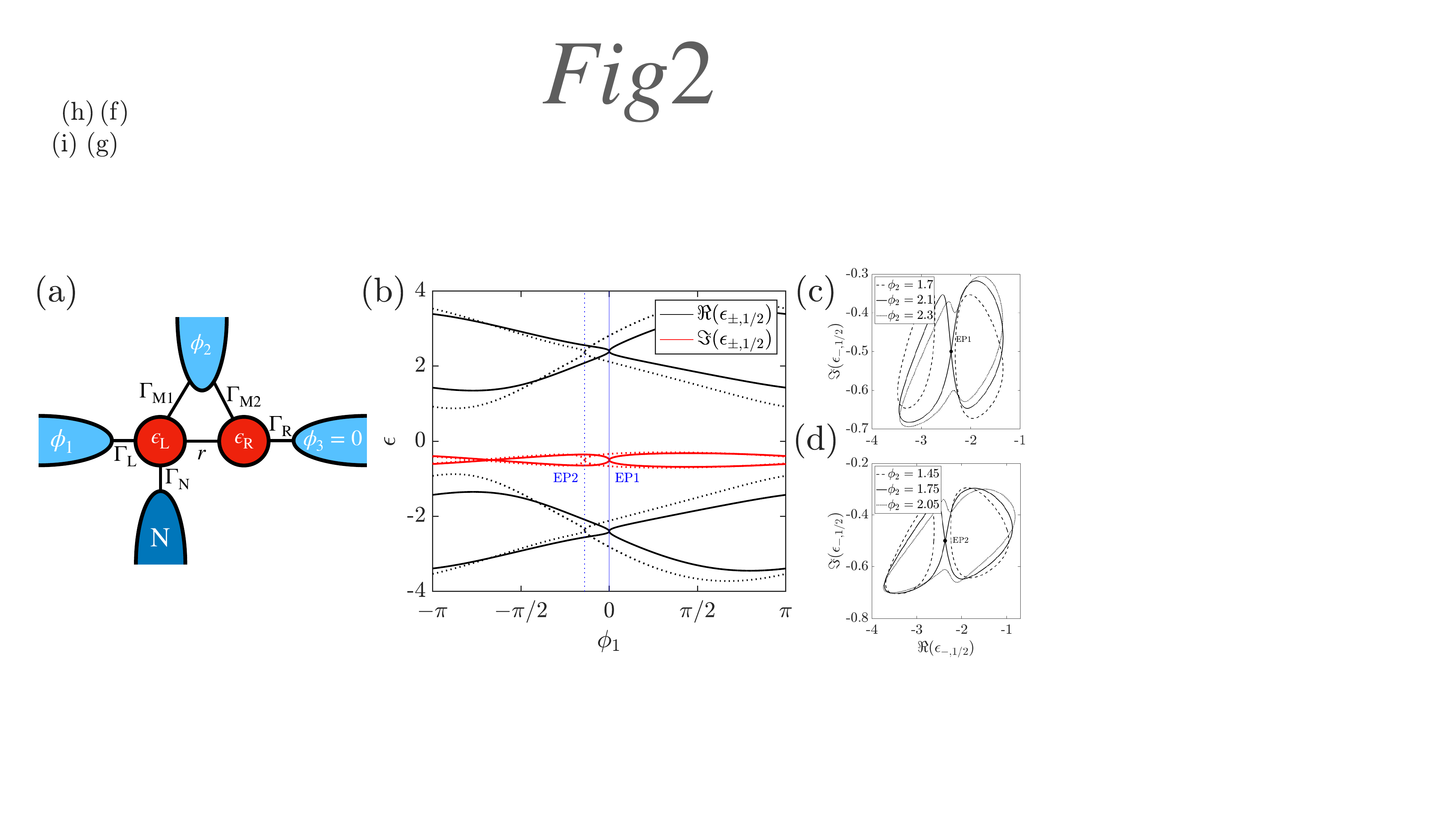}
\caption{(a) Schematic of a double quantum dot three-terminal junction where a normal metal lead adds a non-hermiticity to one of the dots. (b) Complex ABS energies $\epsilon_{\pm,1/2}$ of the effective non-hermitian Hamiltonian depending on the phase $\phi_{1}$ for two parameters configurations. Solid lines correspond to parameters $\Gamma_{\rm L/R/M1/M2} = \Gamma_{\rm N} = 1$, $\epsilon_{1/2} = 0$, $r = 2$ and $\phi_2 = 2.1$ whereas dotted lines correspond to $\Gamma_{\rm L} = 1.4$, $\Gamma_{\rm R}= 1.1$, $\Gamma_{\rm M1} = 1.0$, $\Gamma_{\rm M2} = 0.5$, $\Gamma_{\rm N} = 1$, $\epsilon_1 = 0.1$, $\epsilon_2 = 0.3$, $r = 2$ and $\phi_2 = 1.75$. The positions of the EPs are indicated by the solid and dashed vertical line. (c) The energy spectrum $\epsilon_{-,1/2}(\phi_1)$ for fixed values of $\phi_2$ indicated in the legend. The system goes from two gapped bands, to exhibiting an EP where both eigenvalues coalesce, to a winding where the two bands form a combined loop. (d) The same as in (c) except for the other parameter configuration.}
\label{Fig2}
\end{figure*}

\emph{Stable exceptional points in three-terminal superconducting junctions}.-- To stabilize the EPs we introduce another SC terminal, like it is shown in Fig.~\ref{Fig2}(a). 
It should be mentioned that the multiterminal Josephson junction utilized here was used to explain recent experimental findings on ABS hybridization in Refs.~\cite{Coraiola2023b,Ohnmacht2024}. Similarly to before, one can express the effective Hamiltonian
\begin{widetext}
\begin{align}
H_{\text{eff}}=\bm{d}^\dagger\begin{pmatrix}
\epsilon_\text{L}\tau_3+ (\Gamma_{\rm L}e^{i\phi_1\tau_3}+\Gamma_{\rm M1} e^{i\phi_2\tau_3})\tau_1-i\Gamma_{\text{N}}\tau_0 & r\tau_3 + \Gamma_{\rm M12} e^{i\phi_2\tau_3}\tau_1\\ 
r\tau_3 + \Gamma_{\rm M12} e^{i\phi_2\tau_3}\tau_1 & \epsilon_{\rm R}\tau_3 + (\Gamma_{ \rm M2}e^{i\phi_2\tau_3} + \Gamma_{\rm R})\tau_1
\end{pmatrix}\bm{d}\label{eq:eff2},
\end{align}
\end{widetext}
with the effective couplings $\Gamma_i $ ($i = $ N, R,  M1, M2, L), $\Gamma_{\rm M12} \leq \sqrt{\Gamma_{\rm M1} \Gamma_{\rm M2}}$, the SC phases $\phi_{1,2}$ and the same Nambu spionor from Eq.~\ref{eq:eff}. The value of $\Gamma_{\rm M12}$ depends on the geometry and coherence length in the superconductor \cite{Coraiola2023b,Hofstetter2009,PhysRevLett.104.026801,PhysRevB.63.165314}. We choose $\Gamma_{\rm M12}= \sqrt{\Gamma_{\rm M1} \Gamma_{\rm M2}}$ noting that the following considerations also hold for $\Gamma_{\rm M12} = 0$. This Hamiltonian yields four complex ABSs energies as before. In the following, we treat the system as a 1-dimensional system with the pseudomomentum $\phi_1$ and we fix the phase $\phi_2$ as a control parameter. In Fig.~\ref{Fig2}(b) we show the eigenvalues of the effective Hamiltonian as a function of $\phi_1$ fine tuned to the EP, where black corresponds to the real and red to the imaginary part (solid lines). Another spectrum with varied parameters is shown in Fig.~\ref{Fig2}(b) by the dotted lines. The EP is protected from slight variation of parameters and changes its location which is indicated by the dashed vertical blue lines. This protection can be explained by the existence of a point gap in this system. In  Fig.~\ref{Fig2}(c,d) we show the spectrum of the lower ABSs $\epsilon_{-,1/2}(\phi_1)$ for both parameter configurations used for (b). Observe that in both cases, the ABSs are separated for the lowest value of the control parameter $\phi_2$, but upon increasing it, the ABSs start to touch at one point which corresponds to the EP (indicated by EP1/EP2 respectively). Increasing the control parameter further results in the two ABSs forming a loop and a spectral point gap. This defines a topological winding number. 
Noting that PHS is broken when one EP is shifted towards zero energy and TRS is broken due to fixing $\phi_2$, the system belongs to AZ class A. However, it is not straightforward to classify the gapped system in this case as the parameter $\phi_1$ corresponds to a pseudo-momentum making the system 1-dimensional but not in the conventional sense. Rather, the system has a synthetic dimension of $d_x = 1$ and a spatial dimension of $d_k = 0$ resulting in the classification parameter to be $\delta = (d_k-d_x)\ {\rm mod}\  8 = 7$ \cite{PhysRevB.82.115120, PhysRevB.90.020501}. Considering the point gap and class A, we find a topological invariant of the $\mathbb{Z}$-type corresponding to the winding number of the loop observed in Fig.~\ref{Fig2}(c,d). Note that lowering the coupling $\Gamma_{\rm N}$ results in vanishing EPs, meaning there is a certain non-hermiticity threshold.

\begin{figure*}
\includegraphics[width=1\textwidth]{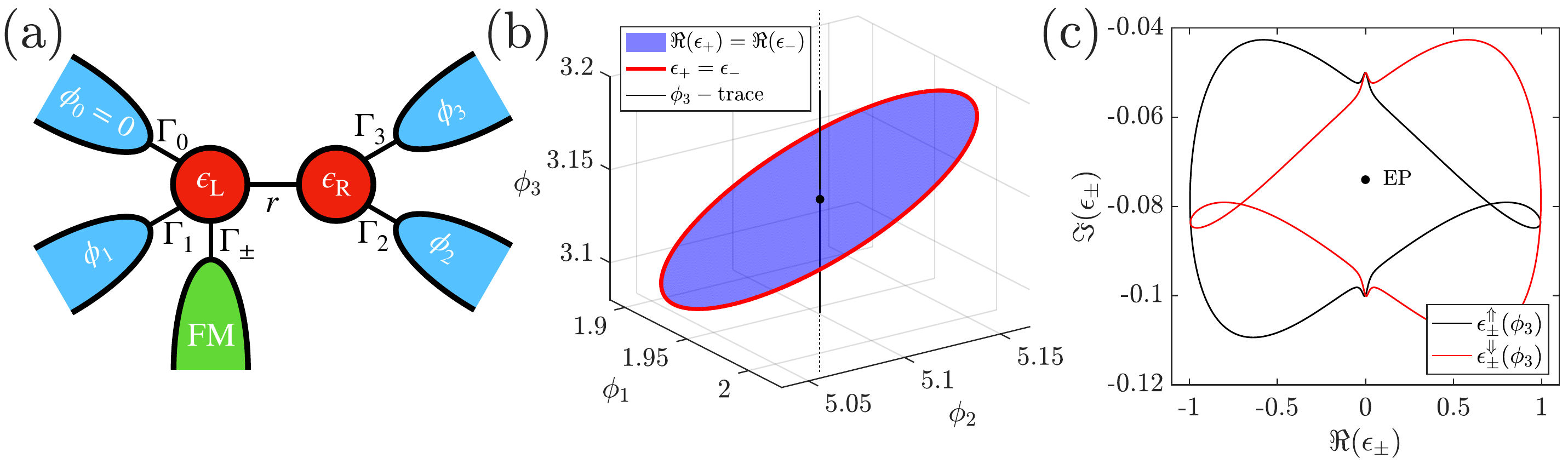}
\caption{(a) Schematic of a double quantum dot four-terminal junction where one dot is coupled to a ferromagnet inducing spin-rotation symmetry breaking and non-hermiticity. (b) An exceptional ring where the real-part of the eigenenergies $\epsilon_{\pm}^{\Uparrow,\Downarrow}$ is degenerate inside the ring (marked by blue) whereas the eigenvalues coalesce on the ring (marked by red). (Parameters chosen: $\Gamma_{0,1,2,3} = 1$, $\Gamma_{+}  = 0.15$, $\Gamma_- = 0.05$, $\epsilon_{1/2} = 1$, $r = 1.5$). The line puncturing the exceptional ring along the $\phi_3$-direction shows the parameters used for the sweep in panel (c). (c) Energy spectrum $\epsilon_{\pm}^\Uparrow(\phi_3)$ and $\epsilon_{\pm}^\Downarrow(\phi_3)$ for the eigenenergies closest to zero (real) energy of the effective non-hermitian Hamiltonian for the line trace in (b) with phases $\phi_1 =1.96 $ and $ \phi_2 =5.1 $.}
\label{Fig3}
\end{figure*}

\emph{Stable exceptional rings arising from hermitian topology in four-terminal superconducting junctions}.-- Stable EPs can also be engineered using Weyl nodes inherent to the hermitian system. For this case, we use another 2-dot model which was recently used in Ref.~\cite{10.21468/SciPostPhys.15.5.214} and is portrayed in Fig.~\ref{Fig3}(a).
In the case of non-zero energies $\epsilon_{\rm L} = \epsilon_{\rm R}\neq 0 $, the two negative (positive) energy ABSs are gapped and Weyl nodes appear for a broad parameter range. In different settings as shown in Ref.~\cite{bergholtz2021exceptional, PhysRevLett.123.066405}, it is expected that a Weyl node in a hermitian system is transformed to an exceptional ring. However, in our case an additional non-hermitian coupling induced by a normal metal does not lead to the generation of an exceptional ring at zero energy. 

To obtain an exceptional ring, one needs to break the spin-rotation symmetry, which is possible by introducing a non-hermitian coupling induced by a ferromagnet, which is spin-polarized, resulting in the following effective Hamiltonian
\begin{widetext}
\begin{align}
H_{\text{eff}}=\frac{1}{2}\bm{d}_{\rm D}^\dagger \Bigg \{ \sigma_0 \otimes \begin{pmatrix}
\epsilon_{\rm L} \tau_3 + (\Gamma_0 + \Gamma_1e^{i\phi_1\tau_3})\tau_1 - i\Gamma_+\tau_0 & r\tau_3\\
r\tau_3 & \epsilon_{\rm R} \tau_3 + (\Gamma_2 e^{i\phi_2\tau_3} + \Gamma_3e^{i\phi_3\tau_3})\tau_1 
\end{pmatrix}+ \sigma_3 \otimes \begin{pmatrix} -i\Gamma_- \tau_3 & 0 \\ 0 & 0  \end{pmatrix} \Bigg \} \bm{d}_{\rm D},
\label{eq:eff3}
\end{align}
\end{widetext}
with the Spin-Nambu spinor $\bm{d}_{\rm D}^\dagger= (d^\Uparrow,d^\Downarrow) = (d_{L\uparrow}^\dagger,d_{L\downarrow},d_{R\uparrow}^\dagger, d_{R\downarrow},d_{L\downarrow}^\dagger,-d_{L\uparrow},d_{R\downarrow}^\dagger, -d_{R\uparrow})$, the SC phases $\phi_{1,2,3}$, the effective couplings $\Gamma_{0,1,2,3}$, the Pauli matrices $\sigma_i$/$\tau_i$ (for $i = 0,1,2,3$) in Spin- and Nambu-space respectively and a spin-dependent coupling $\Gamma_\pm =  \Gamma_{\mathrm{N}} (N_0^\uparrow \pm N_0^\downarrow)/2 $ from spin-dependent densities of states at the Fermi level in the ferromagnet. The spectrum encompasses eight ABSs, where we focus on the four which are close to zero (real) energy. Additionally, the Hamiltonian is block-diagonal, resulting in two different spin-species, which we label by $\epsilon_{\pm}^\Uparrow/\epsilon_{\pm}^\Downarrow$ for the upper/lower block. We find that the original Weyl points transform into an exceptional ring, like it is shown in  Fig.~\ref{Fig3}(b). The blue shaded area corresponds to regions in the phase space where the real parts of the ABSs are degenerate (similar to Weyl disk \cite{PhysRevB.98.241105, PhysRevB.105.235437}) whereas the red ring corresponds to the exceptional ring where the eigenvalues coalesce. To analyze the topology, the system can be treated as either an effective 1D or 2D bulk system, where we obtain either a point gap when a 1D line punctures the exceptional ring or a line gap when a plane crosses the Weyl disk. For the line going along the $\phi_3$-direction in Fig.~\ref{Fig3}(b), the spectrum of the four ABSs $\epsilon_{\pm}^{\Uparrow/\Downarrow}(\phi_3)$ is shown in Fig.~\ref{Fig3}(c). It is seen that, for each spin-block, the ABSs encircle the EP (resulting from the original Weyl point) relating to a point gap and thus a topological invariant. Note, that an arbitrarily small non-hermiticity already leads to a point gap in this system, because of the topological protection of the Weyl node.

The question arises, why spin-rotation symmetry has to be broken in order to get a winding. In a spin-rotation symmetric system ABSs come in particle-hole symmetric pairs inside the same spin-block. However, particle-hole symmetric states are not able to form a winding at zero energy because their spectrums (along one phase) exhibit opposite orientation. If spin-rotation symmetry is broken, the particle-hole symmetric partners of each ABS are in the other spin-block as seen in Fig.~\ref{Fig3}(c). Thus, two ABSs in one spin-block can have the same orientation and can wind around zero energy to encircle an EP. This holds in a more general sense. Namely, in the systems analyzed in Fig.~\ref{Fig1} and Fig.~\ref{Fig2}, we never find EPs at zero energy. The reason is that spin-rotation symmetry prevents the states from forming loops at zero energy. As shown in the SM in Ref.~\cite{SM}, by introducing spin-rotation symmetry breaking, EPs can appear at zero energy in the these systems as well.
Further insight can be obtained by considering the 38-fold way \cite{kawabata2019symmetry}. The spin-rotation symmetric Hamiltonians can be decomposed in blocks of  Hamiltonians $H_i$ which obey PHS$^\dagger$ of Class C$^\dagger$
\begin{equation}
    U_{\rm C} H_{i}^*(\phi)U_{\rm C}^{-1} = -H_{i}(\phi)
\end{equation}
for $U_{\rm C} = i \tau_y$ and $U_{\rm C}U_{\rm C}^* = -1$. Fixing a phase difference generally breaks TRS of the junction. For a system of synthetic dimension $d_x = 1$ (we fix all but one phase difference) and therefore classification parameter $\delta = 7$, a topological phase is prohibited for class $C^\dagger$ which explains why we don't find an exceptional ring when the non-hermiticity is induced by normal metals. In addition, this explains why we also do not find EPs at zero energy for the system analyzed in Fig.~\ref{Fig2}. 

When the spin-rotation symmetry is broken from SU(2) to U(1) due to a coupling to a ferromagnet, each spin-block breaks PHS of class C$^\dagger$ resulting in each block belonging to class A \cite{RevModPhys.88.035005}, which in fact allows for a nontrivial invariant $\mathbb{Z}$ for a point gap which can result in topologically protected EPs at zero energy.

In the system analyzed in Fig.~\ref{Fig2}, spin-orbit coupling in the form of spin-dependent hopping \cite{Fernández-Fernández_2023, PhysRevLett.88.047903} can induce spin-splitting of the ABSs as shown in Ref.~\cite{SM}. This results in windings establishing around zero energy because of the broken spin-rotation symmetry. This can be especially useful considering that spin-splitting was recently observed in Ref.~\cite{PhysRevX.14.031024} in systems that are described by models similar to the ones used here.

\emph{Discussion.}-- In summary, we show that EPs and non-hermitian topology in different synthetic dimensions can be realized in multiterminal Josephson junctions. We showed that in two-terminal junctions, EPs are not stable and thus not suitable for experimental identification. We also show that multiterminal junctions can host topologically protected points gaps in either three- or four-terminal junctions at (non-) zero energies depending on the type of coupling. Furthermore, we showed that Weyl nodes only transform into exceptional rings in the presence of broken spin-rotation symmetry by introducing ferromagnetic couplings or spin-orbit coupling. The classification scheme as well as the analysis of the complex energy spectrum to find topological signatures like point gaps can in principle be applied to arbitrary systems containing superconductors that combine spatial dimensions as well as synthetic dimensions like 1D chains, planar Josephson junctions or hybrid systems containing ferromagnets. In that sense, topological classification and spectral analysis can clarify which types of experimental setups yield stable exceptional points.

In conclusion, our work combines the physics of engineered multiterminal Josephson junctions with non-hermitian physics. The observation of non-hermitian topology, which can be engineered in different synthetic dimensions, further improves the understanding of non-hermitian systems and allows the utilization of non-hermitian topological systems in quantum transport.

D.C.O. and W.B. acknowledge support by the Deutsche Forschungsgemeinschaft (DFG; German Research Foundation) via SFB 1432 (Project No. 425217212).

\bibliography{references.bib}
\end{document}